# Observation of femto-joule optical bistability involving Fano resonances in high-$Q/V_m$ silicon photonic crystal nanocavities

Xiaodong Yang[1,*], Chad Husko[1], Mingbin Yu[2], and Dim-Lee Kwong[2] and Chee Wei Wong[1,*]

[1]*Optical Nanostructures Laboratory, Columbia University, New York, NY 10027*

[2]*The Institute of Microelectronics, 11 Science Park Road, Singapore Science Park II, Singapore, Singapore 117685*

We observe experimentally optical bistability enhanced through Fano interferences in high-$Q$ localized silicon photonic crystal resonances ($Q$ ~ 30,000 and modal volume ~ 0.98 cubic wavelengths). This phenomenon is analyzed through nonlinear coupled-mode formalism, including the interplay of $\chi^{(3)}$ effects such as two-photon absorption and related free-carrier dynamics, and optical Kerr as well as thermal effects and linear losses. Our experimental and theoretical results demonstrate Fano-resonance based bistable states with switching thresholds of 185 μW and 4.5 fJ internally stored cavity energy (~ 540 fJ consumed energy) in silicon for scalable optical buffering and logic.

* Electronic mail: xy2103@columbia.edu, cww2104@columbia.edu

Two-dimensional photonic crystal (2D PhC) slabs confine light by Bragg reflection in-plane and total internal reflection in the third dimension. Introduction of point and line defects into 2D PhC slabs create localized resonant cavities and PhC waveguides respectively, with ab initio arbitrary dispersion control. Such defect cavities in high-index contrast materials possess strong confinement with subwavelength modal volumes ($V_m$) at ~ $(\lambda/n)^3$, corresponding to high field intensities per photon for increased nonlinear interaction. Moreover, cavities with remarkable high quality factors ($Q$) [1, 2] have been achieved recently, now permitting nanosecond photon lifetimes for enhanced light-matter interactions. The strong localization and long photon lifetimes in these high-$Q/V_m$ photonic crystal nanocavities point to enhanced nonlinear optical physics, such as Lorentzian-cavity-based bistability [3-6], Raman lasing [7, 8] and cavity QED [9] in silicon photonics.

The interference of a discrete energy state with a continuum can give rise to sharp and asymmetric lineshapes, referred to Fano resonances [10, 11]. Compared to a Lorentzian resonance, these lineshapes arise from temporal pathways that involve direct and indirect (for example a resonance) transmission, with a reduced frequency shift required for nonlinear switching due to its sharper lineshape. If the indirect pathway can be further strongly localized (such as in a high-$Q/V_m$ 3D cavity instead of a 1D Fabry-Perot cavity), the nonlinear characteristic switching thresholds can be further reduced. Optical bistability involving Fano resonances due to Kerr effect in photonic crystal cavities has been theoretically studied based on Green's function solution of Maxwell's equations [12]. Fano resonances have also been studied by transfer matrix technique [11, 13], and coupled-mode equations [14].

In this Letter, we present our measurements on Fano-based optical bistability as well as a temporal nonlinear coupled-mode framework for numerical analysis. Figure 1(a) shows a schematic of the theoretical model. A waveguide with two partially reflecting elements is side-coupled to a cavity. $a$ is the amplitude of the cavity mode which is normalized to represent the energy of the cavity mode $U = |a|^2$ and $s$ is the amplitude of the waveguide mode which is normalized to represent the power of the waveguide mode $P = |s|^2$. With the coupled-mode formalism [15, 16], the dynamic equation for the amplitude $a(t)$ of the resonance mode is [14, 17]

$$\frac{da}{dt}=\left(-\frac{1}{2\tau_{total}}+i(\omega_0+\Delta\omega-\omega_{wg})\right)a+\kappa s_{a1+}+\kappa s_{a2+} \quad (1)$$

As shown in Figure 1(a), $s_{a1-}=\exp(-i\phi)s_{a2+}+\kappa a$ and $s_{a2-}=\exp(-i\phi)s_{a1+}+\kappa a$. $\phi=\omega_{wg}n_{eff}L/c$ is the phase shift. $\kappa$ is the coupling coefficient between the waveguide mode $s(t)$ and $a(t)$, and $\kappa=i\exp(-i\phi/2)/\sqrt{2\tau_{in}}$ [17]. For a lossy partially reflecting element with the amplitude reflectivity $r$ and transmissivity $t$, ($r^2+t^2\le 1$) [15]

$$\begin{pmatrix} s_{aj+} \\ s_{aj-} \end{pmatrix}=\frac{1}{it}\begin{pmatrix} -(r^2+t^2) & -r \\ r & 1 \end{pmatrix}\begin{pmatrix} s_{j+} \\ s_{j-} \end{pmatrix},\ j=1,2 \quad (2)$$

In equation (1), the total loss rate for the resonance mode $1/\tau_{total}$ [3, 5, 7] is

$$1/\tau_{total}=1/\tau_{in}+1/\tau_v+1/\tau_{lin}+1/\tau_{TPA}+1/\tau_{FCA} \quad (3)$$

where $1/\tau_{in}$ and $1/\tau_v$ are the loss rates into waveguide (in-plane) and into freespace (vertical), and $1/\tau_{in/v}=\omega/Q_{in/v}$, $1/\tau=1/\tau_{in}+1/\tau_v$. The linear material absorption $1/\tau_{lin}$ is assumed small since operation is within the bandgap of the silicon material. $1/\tau_{TPA}$ and $1/\tau_{FCA}$ are the loss rates due to two-photon absorption (TPA) and free-carrier absorption (FCA) respectively. The $\Delta\omega$ detuning of the cavity resonance from $\omega_0$ is modeled due to the Kerr effect, free-carrier dispersion (FCD), and thermal dispersion effects under first-order perturbation. $(\omega_0+\Delta\omega)$ is the shifted resonant frequency of the cavity and $\omega_{wg}$ is the input light frequency in the waveguide. With the modeled TPA generated carrier dynamics and thermal transients due to total absorbed optical power (Eq.(31) and Eq.(42) in Ref.7), the coupled nonlinear dynamical behavior of the Fano optical system is numerically integrated.

The optical system consisting of a photonic crystal waveguide side coupled to a high-$Q/V_m$ nanocavity with five linearly aligned missing air holes (*L5*) in an air-bridge triangular lattice photonic crystal slab with thickness of 0.6*a* and the radius of air holes is 0.29*a*, where the lattice period *a* = 420 nm, as shown in Figure 1(b). The shift $S_1$ of two air-holes at cavity edge is 0.02*a* to tune the radiation mode pattern for increasing the *Q* factors. The waveguide-to-cavity separation is five layers of holes. The index contrast at the waveguide input and output facets act as partially reflecting elements with distance *L* of around 1.9 mm to form a Fabry-Perot resonator and perturb the phase of waveguide mode. Figure 1(c) shows the $E_y$ field of the resonance mode mid-slab from 3D FDTD simulations.

The devices were fabricated with the standard integrated circuit techniques in a silicon-on-insulator substrate. A polarization controller and a lensed fiber are used to couple transverse-electric polarization light from tunable laser source into the waveguide. A second lensed fiber collects the transmission from the waveguide output that is sent to the photodetector and lock-in amplifier. The input power coupled to the waveguide is estimated from the transmitted power through the input lensed fiber, waveguide and the output lensed fiber [5]. The total transmission loss of the whole system is around 24.8 dB at wavelength of 1555 nm. At low input power of 20 μW, the measured resonant wavelength $\lambda_0$ is 1556.805 nm. To measure the $Q$ factor, the vertical radiation from the top of only *L5* nanocavity is collected by a 40X objective lens and a 4X telescope through an iris (spatial filter), which will isolate the cavity region only so that there is no any other influence other than the radiation of the cavity mode. The estimated $Q$, based on the full-width at half maximum (FWHM) $\Delta\lambda$ of 52 pm is around 30,000. From 3D FDTD method, the vertical $Q$ factor $Q_v$ is around 100,000 and the in-plane $Q$ factor $Q_{in}$ is around 45,000 so that the total $Q$ factor $Q_{tot} = 1/(1/Q_v+1/Q_{in}) = \sim 31{,}000$.

Figure 2(a) shows the measured transmission spectrum of the waveguide with different input powers. Each transmission shown is repeated over multiple scans. Sharp and asymmetric Fano lineshapes are observed. The spectral lineshapes depend on the position of cavity resonance in a Fabry-Perot background, highlighting Fano interference pathways. Here the spectra show ascending Fano resonances. The Fabry-Perot fringe spacing $d\lambda$ is around 230 pm, which corresponds to the distance between two waveguide facets $d$ = 1.902 mm ($d = \lambda^2/(2*d\lambda*n_{eff})$ and effective index of 2.77 from FDTD simulations). As the input power increases, the Fano lineshapes were red-shifted due to two-photon-absorption induced thermo-optic nonlinearities in silicon [3-5]. Figure 2(b) shows the calculated transmission spectrum from nonlinear coupled-mode model with the input powers used in the experiment. All parameters used in calculation are from either reference papers or FDTD results [7]. When the input power is 1 μW or less, the cavity response is in the linear regime. As the input power increases, the Fano lineshapes were red-shifted.

Figure 3(a) shows the observed hysteresis loop of Fano resonance at red detuning $\delta$ of 22 pm ($\delta/\Delta\lambda$=0.423). For comparison, the inset of Figure 3(a) shows the measured hysteresis loop for Lorentzian resonance at the detuning of 25 pm for a *L5* nanocavity, where the resonance wavelength is $\lambda_0$ = 1535.95 nm, and $Q$ ~ 80,000, the detuning $\delta/\Delta\lambda$ = 2.6 $> \sqrt{3}/2$. The bistable loops of ascending Fano lineshapes are very distinct from Lorentzian lineshapes. Firstly, one suggestive indication is the asymmetry in the hysteresis loop, with sharp increase (gentle decrease) with increasing (decreasing) power for lower (upper) branch, resulting from the asymmetric Fano lineshape. Secondly, for ascending Fano resonances, an important indication is the upward slope (*increase* in transmission) for *increasing* input power for a side-coupled cavity. For a symmetric Lorentzian in a side-coupled drop cavity, a downward slope (or decrease in transmission) should be expected for increasing input power [18-20]. Thirdly, the dip in the transmission (as indicted by the dotted red circle in Figure 3(a)) is another signature of the Fano resonance. This feature is not observable with a symmetric Lorentzian and in fact is an aggregate result of the three self-consistent solutions of the nonlinear Fano system, such as predicted using Green's function method in Ref. 12. Our nonlinear coupled-mode theory framework cannot trace out the individual solutions [12] but show the aggregate behavior, and is in remarkable agreement with our experimental measurements and the Green's function predictions.

The Fano bistable "off" power ($p_{off}$) is estimated at 147 μW and the "on" power ($p_{on}$) at 189 μW for a 22 pm detuning, as shown in Figure 3(a). These threshold powers are determined experimentally from half the total system transmission losses. From the 189 μW (147 μW) $p_{on}$ ($p_{off}$) thresholds, this corresponds to an estimated internally stored cavity energy [3] of 4.5 fJ (1.5 fJ) based on a numerical estimate of waveguide-to-cavity coupling coefficient ($\kappa^2$) of 13.3 GHz. The consumed energy, in terms of definition used in Ref. 19, is ~ 540 fJ (60 fJ) based on the numerical estimated thermal relaxation time of 25 ns and 11.4 % (1.6 %) of input power absorbed by TPA process for "on" ("off") state, although this could be much lower with minimum detuning to observe bistability. The femto-joule level switching in the stored cavity energy is due to the lowered threshold from the sharp Fano interference lineshape, the small mode volume and high-$Q$ photonic crystal cavities. For the 22 pm detuning, the switching intensity contrast ratio is estimated at 8.5 dB (from the regions with sharp discrete bistable "jumps") with a $p_{on}/p_{off}$ ratio of 1.286. Figure 3(b) shows the calculated Fano bistable hysteresis at the detuning of 22 pm from nonlinear coupled-mode theory. The calculated $p_{off}$ and $p_{on}$ thresholds are 151 μW (with the stored cavity energy of 1.5 fJ) and 186 μW (4.5 fJ) respectively, with a switching contrast of 9.3 dB and $p_{on}/p_{off}$ ratio of 1.232, in excellent agreement with the experimental results.

Now we examine parametrically the dependence of the Fano-type bistability against achievable device characteristics, with our developed nonlinear model. Figure 4 summarizes the extensive numerically-calculated effects of normalized-detuning ($\delta/\Delta\lambda$), mirror reflectivity $r$, cavity $Q$, and the position of cavity resonance on the characteristic threshold power $p_{on}$, and switching contrast. A baseline $Q$ of 30,000, a $r$ of 0.5 with 11% mirror loss, a $\lambda_0$ of 1556.805 nm, and a detuning of $\delta/\Delta\lambda$=22pm/52pm=0.423 is used, which correspond to the current experimental parameters and are represented as the red-filled symbols in Figure 4. In Figure 4(a), for both Fano bistability (solid lines) and Lorentzian bistability ($r$=0, dashed lines), the threshold power increases for increasing normalized detuning (further normalized shift of incident laser frequency from the cavity resonance) due to the larger shift in resonance needed for bistable switching. The minimum detuning required for Lorentzian bistability is $\delta/\Delta\lambda$~0.7, which agrees well with the theoretical threshold detuning $\delta/\Delta\lambda = \sqrt{3}/2$ [21]. However, the threshold detuning for Fano bistability to appear is $\delta/\Delta\lambda$~0.3, which is much smaller than Lorentzian case. The threshold power is similar for both cases (around 140 μW) at the threshold detuning. For both cases, the switching contrast *decreases* with increasing detuning due to the reduced contrast in the transmission at the higher input powers needed for the bistable operation. Compared to Lorentzian bistability, Fano bistability has higher switching contrast (9.87 dB at threshold detuning) and it decreases more slowly with increased wavelength detuning. The low threshold detuning and high switching contrast for Fano bistability is due to the sharp and asymmetric Fano lineshapes. The inset of Figure 2(b) plots the transmission spectrums of Fano lineshapes and Lorentzian lineshapes with input power of 1 μW and 230 μW respectively. The sharp transition right before the cavity resonance makes Fano bistability to occur with much lower wavelength detuning, while Lorentzian bistability needs a larger detuning to reach the multivalue transmission regime. For the detuning shown in the figure, the switching contrast of Fano bistability (marked as a red arrow) is higher than Lorentzian bistability (marked as a blue arrow). There is significant difference in term of wavelength detuning required for bistability to occur and the bistable switching contrast between Fano bistability and Lorentzian bistability. This switching contrast can significantly increase when the mirror reflectivity $r$ increases from 0.35 to 0.8 (at detuning of $\delta/\Delta\lambda$ = 0.423) at the expense of increasing $p_{on}$ (Figure 4(b)). The increase in $p_{on}$ is due to higher mirror reflectivity, resulting in lower power coupled into the Fano system. A limit of 0.35 is used because for smaller $r$, a combination of both Lorentzian and ascending Fano resonance starts to appear. For $r$ greater than 0.8, the threshold power will go up even higher.

Figure 4(c) plots the threshold power and the stored cavity energy with different cavity $Q$ factors at detuning of $\delta/\Delta\lambda$=0.423 and $r$=0.5. Note that $p_{on}$ shows a ($1/Q^{1.569}$)-dependence, while the stored cavity energy needed for bistable shows a ($1/Q^{0.689}$)-dependence. For cavity $Q$ factor of half a million, the Fano threshold power is estimated at 2.4 μW, which corresponds to the stored cavity energy of 0.55 fJ. This stored cavity energy is much lower than a Fano resonance with cavity $Q$ of 30,000 (4.5 fJ). The comparison between Fano bistability and Lorentzian bistability at threshold detuning shows that Fano system has similar threshold power and stored cavity energy as Lorentzian system. We also note that direct comparisons between an ascending Fano-type bistability and a Lorentzian-type bistability in term of threshold power are difficult because the Fano system has different threshold detuning from Lorentzian system and Fano system also depends on additional parameters such as mirror reflectivity $r$.

Figure 4(d) illustrates the influence of different position of cavity resonance $\lambda$ relative to experimental $\lambda_0$ within the half period of Fabry-Perot background $d\lambda/2$ with $Q$=30,000, $r$=0.5, and $\delta/\Delta\lambda$=0.423, where the limits of $(\lambda-\lambda_0)/(d\lambda/2)$ are from -0.8 to 0.2 for ascending Fano resonances. These limits are chosen because they cover the region where the ascending Fano resonances are dominant over the Lorentzian or the descending Fano resonances. For ascending Fano resonances at detuning of $\delta/\Delta\lambda$=0.423, both threshold power and switching contrast increase as the cavity resonance $\lambda$ shifts from the Fabry-Perot background maximum to its minimum. The switching contrast has a maximum at a region close to the minimum Fabry-Perot background, illustrating an interesting trade-off when selecting an optimum set of Fano-type bistable operating parameters. The Fano resonance has little control in the current configuration, which depends on the Fabry-Perot cavity formed by two waveguide facets. For different devices, either ascending or descending Fano resonance can be obtained, depending on the position of cavity resonance in the Fabry-Perot background. In the future, integrated partially reflecting mirrors embedded inside photonic crystal waveguide can be adopted [22]. By tuning the distance between the air holes, the Fabry-Perot background can be tuned to achieve the designed Fano lineshapes.

In this work we demonstrate experimentally all-optical bistability arising from sharp Fano resonances in high-$Q/V_m$ silicon photonic crystal nanocavities. Using the two-photon-absorption induced thermo-optic nonlinearity, an "on"-state threshold of 189 μW and stored cavity energy of 4.5 fJ is observed, and in good agreement with the nonlinear coupled-mode formalism. Although the thermo-optic is slow (on order of μs), other nonlinear mechanisms such as two-photon-absorption induced free-carrier dispersion [3, 4, 6] can remarkably achieve ~ 50 ps switching in silicon. The threshold power can be further reduced to the μW level (or sub-fJ of stored cavity energy) with higher-$Q/V_m$ nanocavities or further optimization of the detuning for reduced threshold and large contrast ratio. Our observations of Fano-type bistability highlight the feasibility of an ultra-low energy and high contrast switching mechanism in monolithic silicon benefiting from the sharp Fano lineshapes, for scalable functionalities such as all-optical switching, memory, and logic for information processing.

This work was partially supported by DARPA and the National Science Foundation (ECCS-0622069). X. Yang acknowledges the support of an Intel Fellowship.

**Figure captions**

Fig. 1. (a) Schematic of optical system including a waveguide side-coupled to a cavity. Two partially reflecting elements are placed in the waveguide. (b) SEM of photonic crystal *L5* point-defect cavity side-coupled to line-defect waveguide. The input and output facets of the high-index-contrast waveguide form the partially reflecting elements. (c) $E_y$–field of the resonance mode mid-slab from 3D FDTD simulations.

Fig. 2. Measured (a) and CMT-calculated (b) transmission spectrum at different input powers, illustrating the asymmetric lineshapes. The side-coupled *L*5 cavity has a total $Q$ of ~ 30,000. The inset of (b) plots the calculated transmission spectrums of Fano lineshapes (red) and Lorentzian lineshapes (blue) with input power of 1 μW (solid line) and 230 μW (dashed line) respectively.

Fig. 3. (a) Measured and (b) CMT-calculated asymmetric hysteresis loops for Fano resonance at a detuning of 22 pm. The red-circled region in panel (a) highlights a dip in transmission with increasing input power, a signature not present in Lorentzian-type resonances, and indicative of nonlinear Fano-type solutions. The inset of (a) shows the measured hysteresis loop for Lorentzian resonance. The arrows depict ascending and descending input powers to the Fano system.

Fig. 4. CMT calculated effects of (a) the wavelength detuning $\delta/\Delta\lambda$. Dashed lines (--) represent Lorentzian bistability which, compared to this particular Fano-type bistability, requires higher detuning to observe bistability and results in lower switching contrast. Effects of (b) mirror reflectivity $r$, (c) cavity $Q$ factor, and (d) the position of cavity resonance on the switching threshold power $p_{on}$ and switching contrast. (The red-filled symbols correspond to the experimental parameters.)

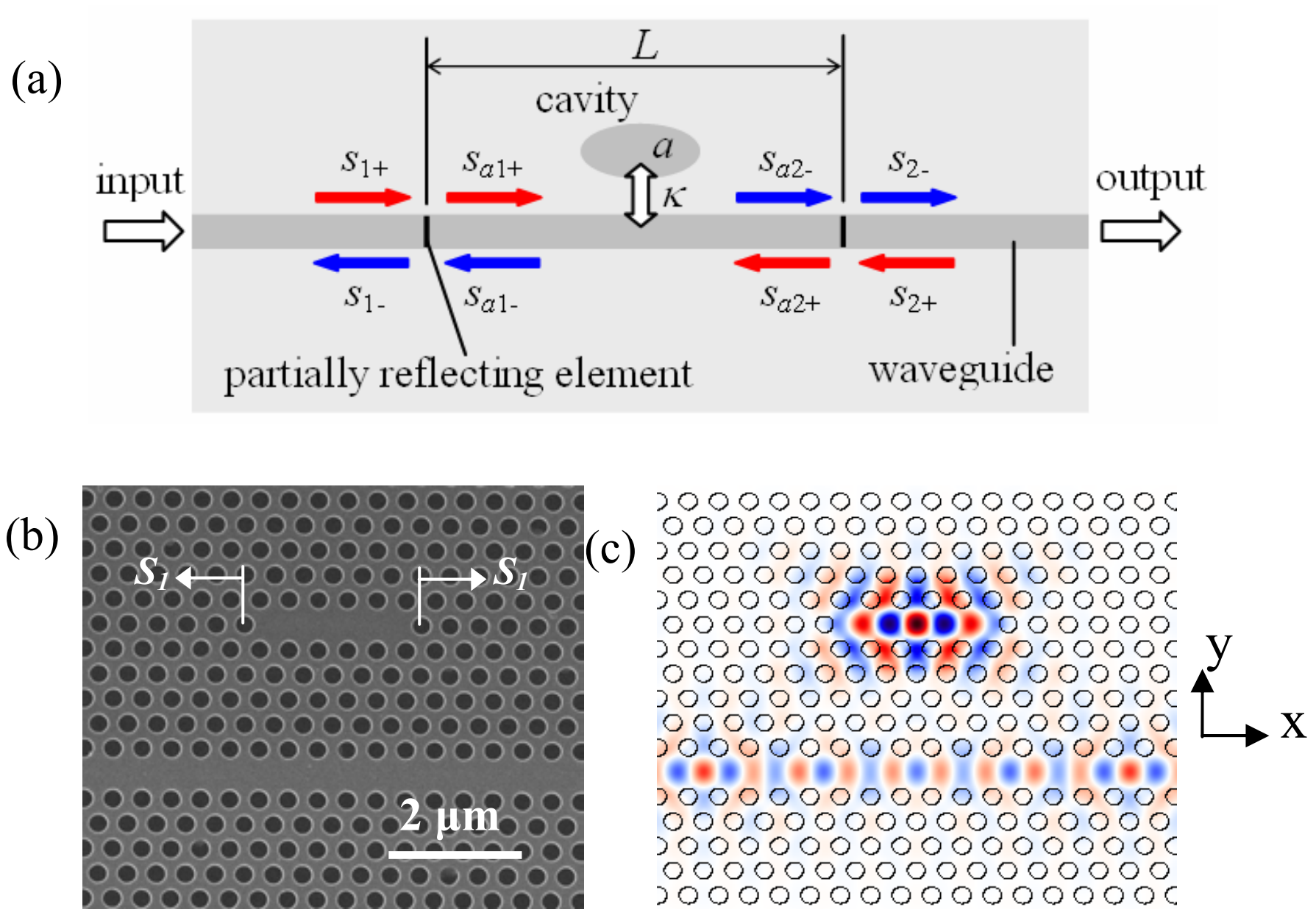

Fig. 1. (a) Schematic of optical system including a waveguide side-coupled to a cavity. Two partially reflecting elements are placed in the waveguide. (b) SEM of photonic crystal *L5* point-defect cavity side-coupled to line-defect waveguide. The input and output facets of the high-index-contrast waveguide form the partially reflecting elements. (c) $E_y$–field of the resonance mode mid-slab from 3D FDTD simulations.

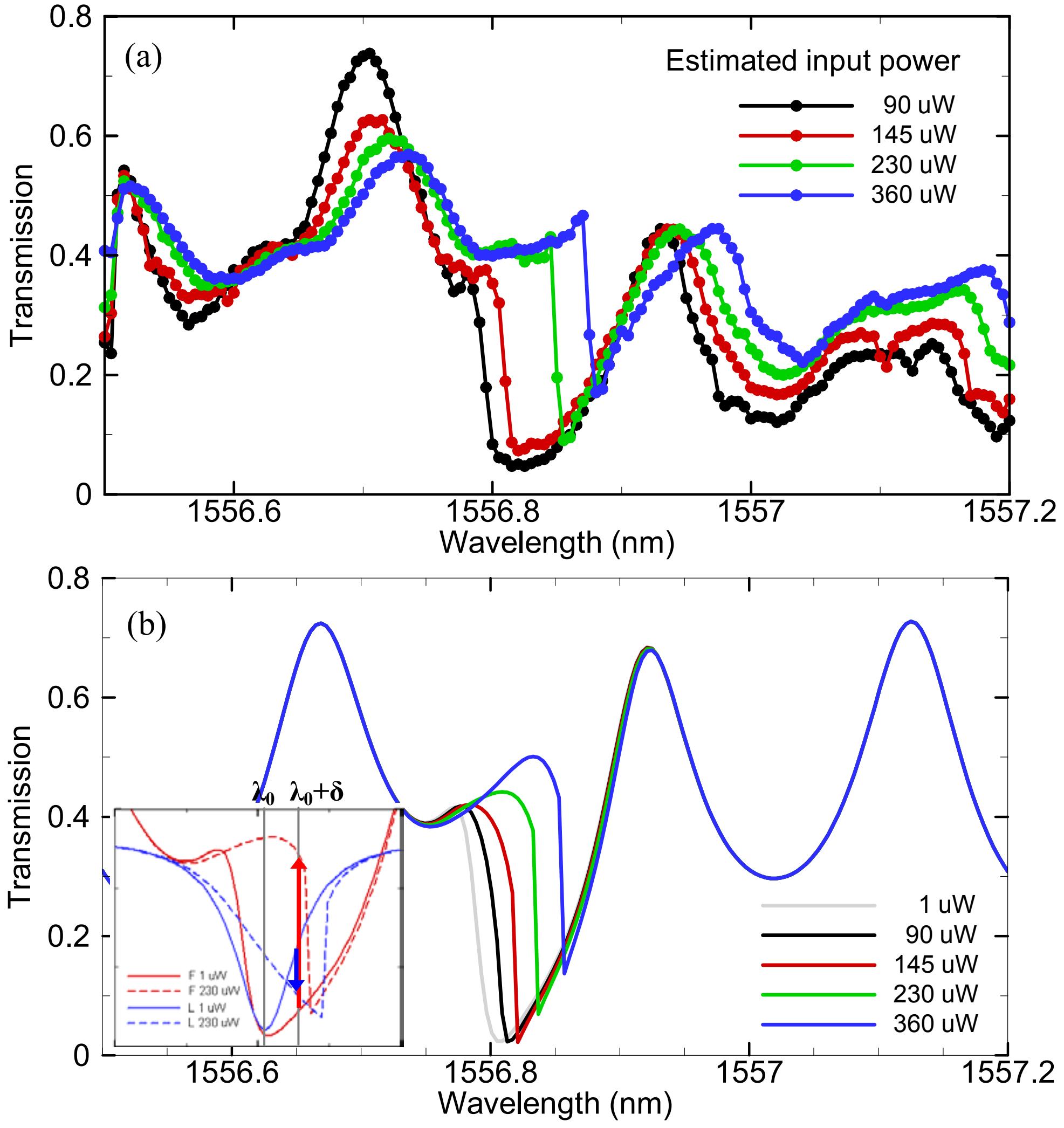

Fig. 2. Measured (a) and CMT-calculated (b) transmission spectrum at different input powers, illustrating the asymmetric lineshapes. The side-coupled *L*5 cavity has a total *Q* of ~ 30,000. The inset of (b) plots the calculated transmission spectrums of Fano lineshapes (red) and Lorentzian lineshapes (blue) with input power of 1 μW (solid line) and 230 μW (dashed line) respectively.

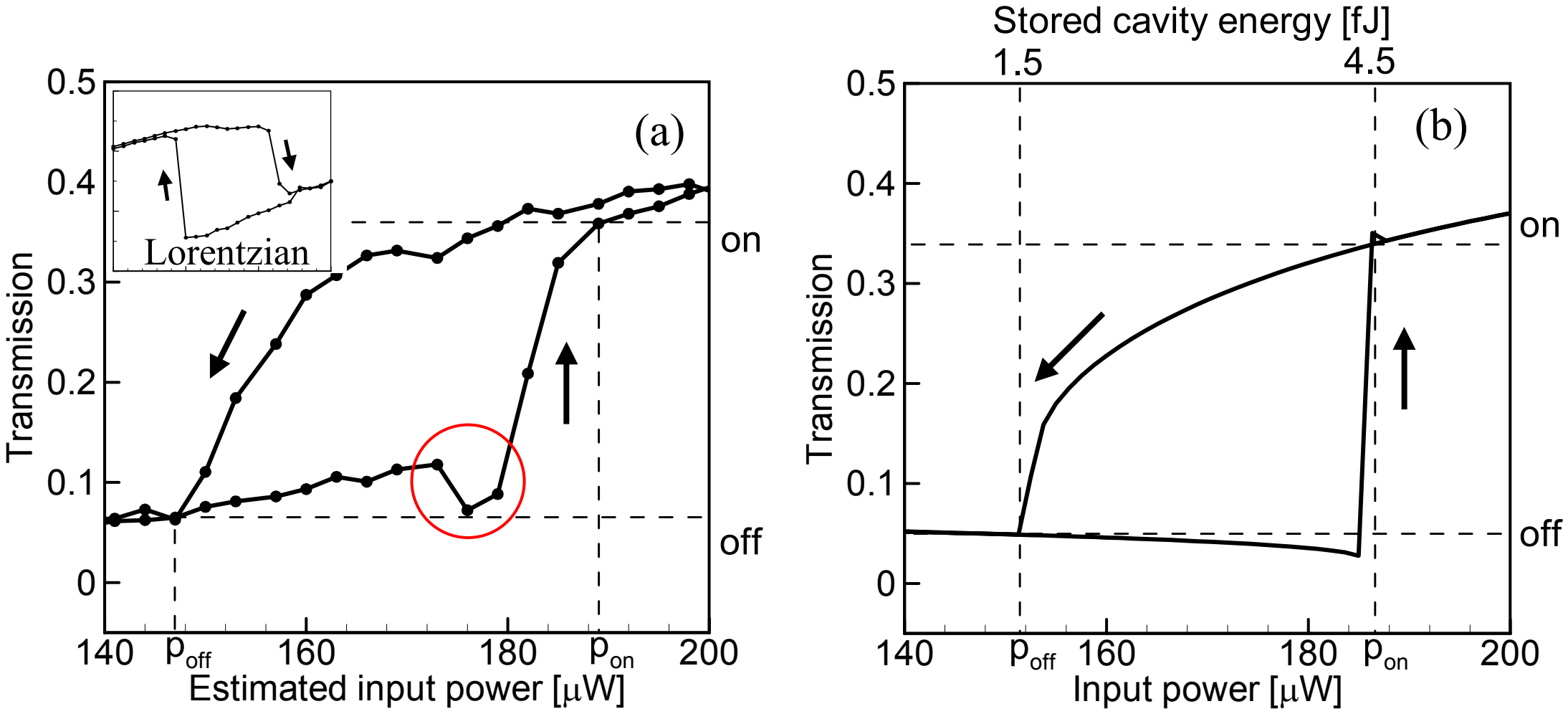

Fig. 3. (a) Measured and (b) CMT-calculated asymmetric hysteresis loops for Fano resonance at a detuning of 22 pm. The red-circled region in panel (a) highlights a dip in transmission with increasing input power, a signature not present in Lorentzian-type resonances, and indicative of nonlinear Fano-type solutions. The inset of (a) shows the measured hysteresis loop for Lorentzian resonance. The arrows depict ascending and descending input powers to the Fano system.

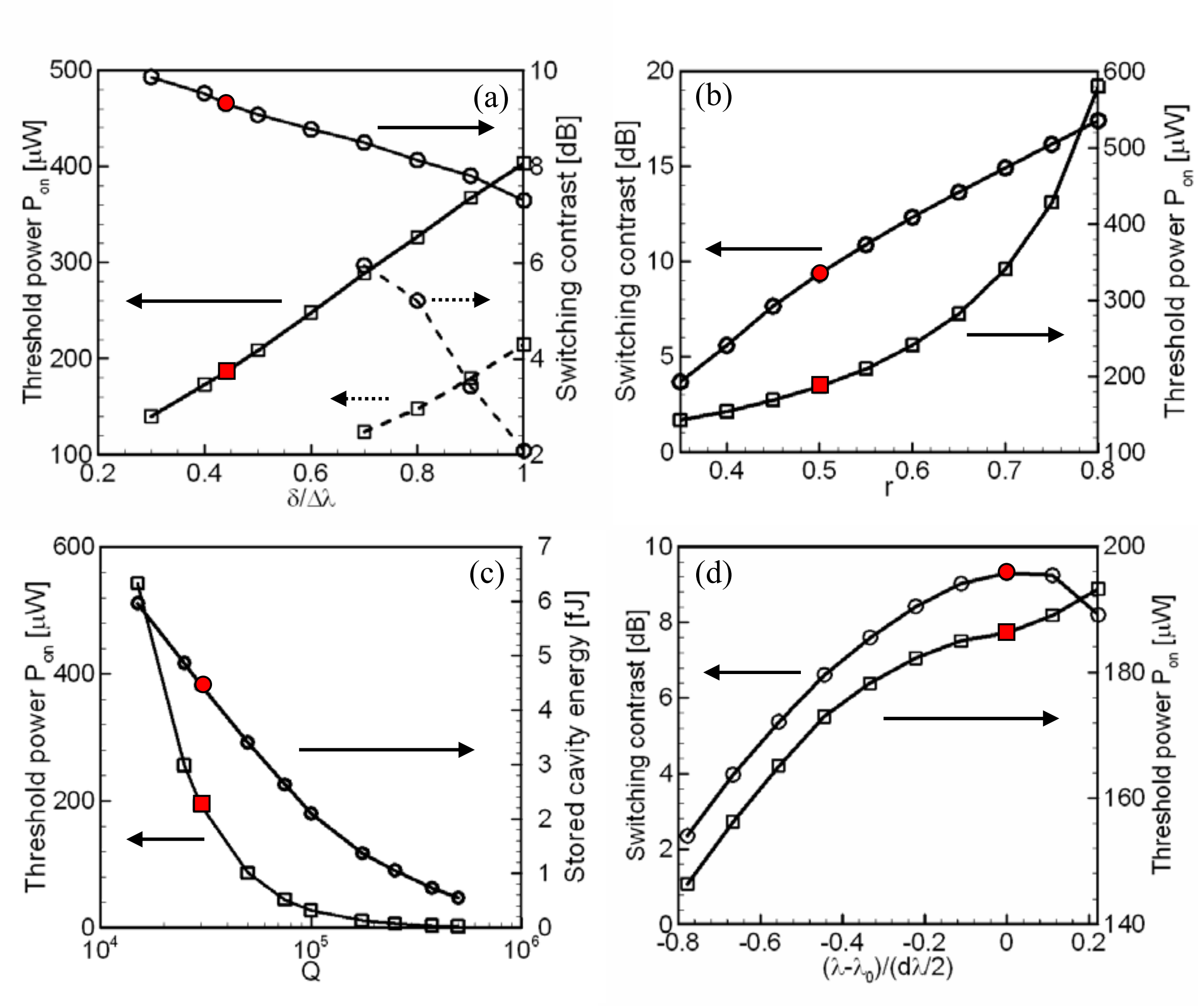

Fig. 4. CMT calculated effects of (a) the wavelength detuning *δ/Δλ*. Dashed lines (--) represent Lorentzian bistability which, compared to this particular Fano-type bistability, requires higher detuning to observe bistability and results in lower switching contrast. Effects of (b) mirror reflectivity *r*, (c) cavity *Q* factor, and (d) the position of cavity resonance on the switching threshold power $p_{on}$ and switching contrast. (The red-filled symbols correspond to the experimental parameters.)